\documentclass[preprint2]{aastex63}
\usepackage{hyperref}
\usepackage{tikz}

\newcommand\aastex{AAS\TeX}

\makeatletter
\global\let\tikz@ensure@dollar@catcode=\relax
\makeatother

\shorttitle{\aastex\ + Overleaf template article}
\begin{document}

\title{Random Forests applied to High Precision Photometry Analysis with Spitzer IRAC.}

\correspondingauthor{Jessica Krick}
\email{jkrick@caltech.edu}

\author{Jessica E. Krick}
\affil{Caltech/IPAC}

\author{Jonathan Fraine}
\affiliation{Space Science Institute}

\author{Jim Ingalls}
\affiliation{Caltech/IPAC}

\author{Sinan Deger}
\affiliation{Caltech/IPAC}



\begin{abstract}

 We present a new method employing machine learning techniques for measuring astrophysical features by correcting systematics in IRAC high precision photometry using Random Forests.  The main systematic in IRAC light curve data is position changes due to unavoidable telescope motions coupled with an intrapixel response function.  We aim to use the large amount of publicly available calibration data for the single pixel used for this type of work (the sweet spot pixel) to make a fast, easy to use, accurate correction to science data.  This correction on calibration data has the advantage of using an independent dataset instead of using the science data on itself, which has the disadvantage of including astrophysical variations.  After focusing on feature engineering and hyperparameter optimization, we show that a boosted random forest model can reduce the data such that we measure the median of ten archival eclipse observations of XO-3b to be $1459 \pm 200$ parts per million.  This is a comparable depth to the average of those in the literature done by seven different methods, however the spread in measurements is 30-100\% larger than those literature values, depending on the reduction method.   We also caution others attempting similar methods to check their results with the fiducial dataset of XO-3b as we were also able to find models providing initially great scores on their internal test datasets but whose results significantly underestimated the eclipse depth of that planet.  

\end{abstract}




\section{Introduction} \label{sec:intro}
We seek photon limited photometry for {\it Spitzer} IRAC data \citep{2004ApJS..154...10F, 2004ApJS..154....1W}.  While this goal is challenging, it is worth pursuing because of the interesting science of atmospheric characterization of exoplanets and brown dwarfs that is enabled by high precision photometry with an infrared instrument in space.  The {\it Spitzer} space telescope operated from 2003 till early 2020 and spent about 30\% of its observing time looking at exoplanets and brown dwarfs.  That is a total of many years of valuable, space-based data.

The main systematic in high precision photometry with {\it Spitzer} IRAC  is the intrapixel response function.  We see variations in the fluxes in light curves due to the combination of under-sampling with small motions of the telescope which move the light around within a single pixel, and since such a large fraction of the light in the PSF falls in that single pixel, the total light curve varies by up to 8\% and 5\%in ch1 and ch2 respectively (for a description of instrument details see \url{https://irsa.ipac.caltech.edu/data/SPITZER/docs/irac/iracinstrumenthandbook/}).  Because motions of the telescope are unavoidable, researchers have developed non-linear analysis methods to  measure astrophysical variations (transits/eclipses/phase variations) in the midst of these correlated signals \citep{2010PASP..122.1341B, 2012ApJ...754..136S, 2013ApJ...765..127F, 2013ApJ...766...95L,2015ApJ...805..132D,2015MNRAS.451..680E, 2016ApJ...820...86M, krick2016}. For a full discussion of comparing these methods side-by-side, see \cite{2016AJ....152...44I}. That paper particularly focused on repeatability and accuracy as tools to asses the variability of the reduction methods.

%This approach revealed that wavelet+ICA (independent component analysis) provided a dramatic improvement over other methods. Notably, Gaussian Processes did not perform significantly better than other methods, and failed at most reproducibility metrics.

Inspired by this work, and the easily available machine learning (ML) tools, we have begun to investigate rigorous ML models to best incorporate subject expertise and smart computer techniques to use big datasets. The goal of this project is to  accurately measure astrophysical variations by generating a model which uses some features of the calibration dataset to predict intrapixel response.   The term 'features' is machine learning language meaning variables that have been measured (or derived from quantities measured).  We can then apply that model to science datasets to predict response values which will reduce the correlated noise in light curves.  The idea here is that for our calibration dataset, we have measured all of the features we can think of, as well as the flux.  This is known as a labelled training set where the label is the flux, which is the quantity we are trying to predict in our science datasets. Importantly, the calibration dataset flux values vary only due to systematics; it is a standard star with a demonstrably flat light curve.  If we can find a model which accurately predicts the flux of the calibration star, then that model will have removed the systematics from our light curves. We will then be able to use that model to accurately measure the astrophysical variations in exoplanet and brown dwarf light curves by removing the systematics.  We assume IRAC systematics are time invariant.

Our approach is to combine all calibration data from years of {\it Spitzer} calibration observations. The advantages of this approach are threefold.  First, ML can handle not only large amounts of data, but also many dimensions.  This allows us to explore effects with position, as has traditionally been done, but also potential correlations with noise pixel, pixel values, background values, etc.  The strength of this machine learning approach is that we can explore the effect of any parameter which we can measure on the intrapixel response function. Secondly, we use an independent calibration dataset to correct for the intrapixel response function.  We choose to use an independent dataset instead of the science data themselves so that astrophysical signal is de-coupled from the noise signal, and we are not inadvertently removing astrophysical signal when we remove noise.  The calibration dataset was specifically designed to be observations of stars which are not variable, and are not planet-hosts, at least to the limit of precision {\it Spitzer} IRAC observations.    If we remove astrophysical signal when removing the systematics, then we have failed.  In order to disentangle the systematics from the signal, we make the assumption that we know the ground truth of the eclipse depths of XO-3b (see Section truth) and use that as the fiducial measurement of the eclipse depth thereby allowing us to know if we have correctly removed the systematics from the light curves. Thirdly, once trained, a ML model is extremely fast at predicting fluxes and will then drastically reduce processing time for astronomers wishing to reduce their light curves.  The increase in speed will also allow us to reduce the entire {\it Spitzer} IRAC  archive of long staring observations in a uniform manner \citep{2018SPIEJK}.

The majority of exoplanet observations with {\it Spitzer} IRAC were done in channel 2 at $4.5\micron$ because the intrapixel response is a smaller effect in channel 2, so we focus only on that channel here.  Additionally, the data challenge which we use as a fiducial was only run for channel 2, so we only have a check on the outcomes of the models in channel 2.  However, a similar method could be used on channel 1 ($3.6\micron$).

This is the first paper in a series where we explore three different machine learning algorithms.  This paper focuses on Random forests which offer a relatively easily interpreted, fast method for incorporating many dimensions while minimizing overfitting.   Overfitting is what happens when a model does too good of a job at fitting all features in a training set but that model then fails to reproduce features in other (non-training) sets.  Overfitting is a common problem for machine learning models which are designed to fit training sets as well as possible.  Future papers will cover a modified nearest neighbors approach (KNN) and neural networks (NN).  KNN is interesting for this problem because it builds on successful algorithms already in the literature. Neural networks are extremely powerful at fitting multidimensional space, but are also more difficult to interpret.  

 Random forests are occasionally used in other astronomical applications for regression \citep{2015ApJ...798..122M, 2019ApJ...882...35V,2019ApJ...884...33G}.  Machine learning work to remove systematics from exoplanet light curves was presented in \cite{2016PASP..128i4503W}.  Those authors use a causal pixel model (CPM) to measure spacecraft systematic effects in Kepler time series data.  This is a data driven approach which assumes any star on the detector can predict another star's variability. To do this, they require two important features of Kepler data that are not possible with a guest observatory like {\it Spitzer}.  First, they require light curves of many stars observed at the same time. Second, they require a lot of out of transit data on the target star.  Because most IRAC exoplanet observations are done in subarray mode ($~0.25$ square arcminutes), we mostly have a single, few hour duration observation of a single target which needs to be corrected for systematics.  Also, even for those observations made in full array mode ($~25$ square arcminutes), there are not anywhere near the number of stars available to Kepler's 116 square degrees. Because Kepler observes each field for greater than or equal to 80 days, each target planet host has much more data than the typical IRAC, of order ten hour duration, observation which can be used to separate systematics from astrophysical variation.  As such, while impressive in its results, the CPM model is not applicable to this project.

In Section \ref{sec:data} we describe the data used for this project.  This includes both  both calibration data from which the feature set is derived, and archival scientific data used for testing the accuracy of the model.  Section \ref{sec:ML} discusses the ML models including optimization techniques and performance.  Specifically we discuss a decision tree model in Section \ref{sec:DT}, a random forest model in Section \ref{sec:RF}, and a boosted random forest model in section \ref{sec:xgboost}.  We attempt to interpret the model itself and what it is telling us about the data and the source of the systematics in section \ref{sec:discussion}.  These are followed by conclusions in Section \ref{sec:conclusion}.   The Appendix includes a more detailed discussion of feature importances and permutation importances as a way of interpreting the models.
%, and individual conditional expectation plots.  

\section{The Data} \label{sec:data}
We describe here both the calibration data used as the labelled training set for our models, and the actual {\it Spitzer} exoplanet data on which we compare the performance of our model with proven reduction techniques in the literature.

\subsection{Calibration Data} \label{sec:pmap_data}
The calibration dataset is presented in detail in \cite{2012SPIE.8442E..1YI}. Briefly, for calibrating channel 2 we use photometry of star BD+67 1044 (also known as NPM1+67.0536; 480mJy at $4.5\micron$).  We use a 0.1s exposure time in channel 2. This is $13\%$ of full well in channel 2.  We employed staring mode (no dither pattern) coupled with commanded position offsets and natural drift of the telescope positions to move the image of the star around on the pixel and fill in the sweet spot.  We define the sweet spot to be the peak of sensitivity of the central pixel at (x,y) = (15.12,15.09) in channel 2. The sweet spot is also the position where data taken using our target acquisition program using the Pointing Calibration and Reference Sensor (PCRS) instrument is most likely to fall (called PCRS Peak-up \url{https://irachpp.spitzer.caltech.edu/page/Obs\%20Planning}).  Calibration data were taken over the course of 6 years from 2010 to 2016, when openings in the schedule allowed.  This means that data was taken at many, random spacecraft orientations and positions. Before centroiding and photometry, the images are averaged together in groups of four. After sigma clipping on centroids, fluxes, flux uncertainties, backgrounds, and background uncertainties there are 787,060 total photometry points in our ch2 dataset.

\subsubsection{Feature Set}
  Our goal is to generate a model which can accurately predict the aperture flux of the calibration star as a function of variables that have been measured, or derived from quantities measured;  features. Note: Absolute flux is not important, but it is our goal to remove relative changes in flux due to, for example, position changes.  The feature list used is designed based on years of experience using this instrument and represents our best guess as to the (non-astrophysical) variables on which flux could depend. We note that the machine learning techniques applied here will not discover new features which impact intrapixel response, and are limited to the training set features.  We start with 20 potential features.

\begin{itemize}
\item X \& Y center positions values calculated by center of light
\item X \& Y center positions uncertainties
%\item XY center position covariance
\item Flux uncertainty
\item Noise Pixels (i.e. Beta-Pixels; see Lewis et al. 2013)
\item X \& Y full-width half-max values
%\item Barycentric Modified Julian Date (time array)
\item Temperature of the detector (T\_cernox)
\item Background Flux in 3-7 pixel annulus
\item Background Flux Uncertainty 
\item Central 9 pixel values (3x3 array of pixel values surrounding and including the pixel which includes the stellar peak flux. ) These are measured in units of surface brightness (MJy/sr)
\end{itemize}

\begin{figure*}

 \centering
  \includegraphics[width=\textwidth]{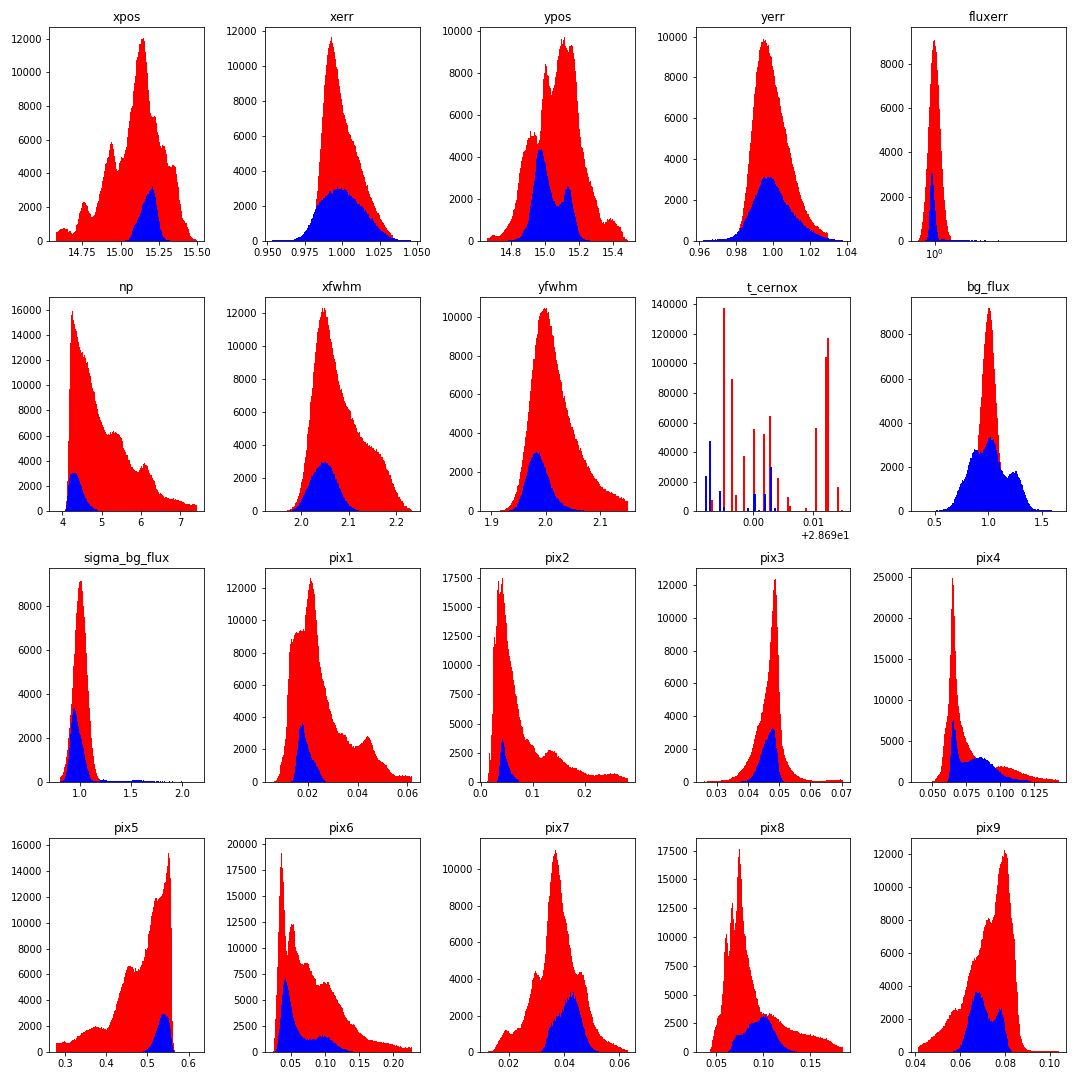}
 \caption{ \label{fig:dist}  {\bf Data Histograms}. Here we show the histograms of each potential data feature for both the calibration dataset(red) and the combined XO-3b science dataset(blue)}
 \end{figure*}

We use center of light centroids measured using the Spitzer Science Center contributed software "Box centroider" to measure the first moment of light in the peak pixel (\url{https://irsa.ipac.caltech.edu/data/SPITZER/docs/irac/calibrationfiles/pixelphase/box_centroider.pro}).  We measure aperture fluxes in a 2.25 pixel radius aperture using the idl code aper.pro with a background measured in an annulus of 3 - 7 pixels. Future applications of the model derived in this work will also then need to use these exact prescriptions.  Future work could look at the effects of changing these prescriptions on the final predictive power of these types of models.

We begin by normalizing the pixel values by the mean of the sum of all nine pixels over all calibration observations, making these features more like a measure of fractional flux.  Additionally we normalize the background fluxes by the mean background level over the whole dataset.  The same is done for the positional errors and the flux and background uncertainties so that these are normalized to a mean value over the whole calibration dataset.  All of this normalization is required to be able to apply the corrections to any dataset with differing flux, background and centroiding error levels, not just the same dataset from which it was made.

We attempted to scale our calibration data using the scikit-learn provided RobustScaler.  Ideally this would have scaled our calibration data to remove the median value and remove outliers using a user specified quantile range, and saved the coefficients.  Then the same scaling could have been applied to the XO-3b data (or any science data).  However, many tests of RobustScaler were performed using different feature sets and or different quantile bounds, but all were found to produce undesirable results in the XO-3b dataset (removed eclipse signal or increased noise).  Since  scalers are not necessary for decision trees, and we have found models which work without scaling, we will continue without them and potentially revisit scaling when working with deep learning models.

We tested models with background subtracting pixel values but found this to be inconsequential in the predictive power of the model.  The background values are only a small fraction of the flux in the central pixels.

We do not remove from our measured fluxes the known extremely small flux degradation \citep[][of order 0.1\% per year in ch1 and 0.05\% per year in ch2]{krick2016} in our fluxes since many observers do not know about this and are likely not to include it in their reductions. The decrease in sensitivity is potentially caused by radiation damage to the optics. Future work could include this in the features; potentially in conjunction with using time as a feature as discussed below.

Although available, we choose not to use time as a feature because machine learning models are  likely poor predictors outside of the range of values that they are trained on, and therefore only times during 2010 - 2016 would be available for study if we included time, which is too constrained for our science goals of being able to use this corrector on any post cryogenic IRAC data.  This would potentially be an interesting avenue to approach in a future paper.  It would be interesting to test within the range of dates that we have calibration data if using time would be a valuable feature by comparing results from models generated both with and without time as a feature.  

Similarly, due to range constraints, we emphasize that the models derived in this paper should be used for data taken at the sweet spot of the detector, and are untested for other pixel positions.

 We do not know which features are important features for removing systematics from IRAC data.  We intentionally test throughout the paper different sets and combinations of features to find a model that can best fit eclipse depths.

\subsection{Exoplanet Data: XO-3b} 
\label{sec:XO3}

We choose the publicly available XO-3b dataset to evaluate performance of our models.  This dataset is extremely valuable in that it has been reduced by seven prominent teams using different methods during an IRAC data reduction challenge.  This dataset was originally published by \citet{2015ApJ...811..122W}.  This XO-3b dataset is the closest thing we have to a dataset where we potentially know the true eclipse depth.  Also, XO-3b was observed ten times in the same way which allows us to examine the reliability of our method. The XO-3b data and data reduction challenge are described in detail in \cite{2016AJ....152...44I}.  Briefly, there are ten individual observations during secondary eclipse of X03-b taken with post-cryogenic IRAC in 2012 and 2013 (PID 90032).  Each observation is 8.5 hours in duration and utilizes a 2s frame time.  XO-3b is a 57.3mJy source observed at about 30\% full well, well within the linear regime of the detector.  This dataset only exists for ch2.  Photometry on this dataset was performed in the same way as for calibration data, including sigma clipping the outliers. In total, all ten eclipse observations combined amount to just under 150,000 datapoints.

We measure the features in the XO-3b dataset in exactly the same way as was done for the calibration dataset.  Histograms of all 20 feature distributions are shown in Figure \ref{fig:dist}.  Larger, red histograms are those for the calibration dataset.  Blue histograms show the XO-3b science comparison dataset.  We employ Freedman-Diaconis binning \citep{Freedman1981} which chooses bin sizes based on the interquartile range and the third root of the number of data points.  This appears to be the most robust technique for examining the relative distributions of these two datasets.

From the distributions shown in Figure \ref{fig:dist}, we note that the XO-3b dataset does indeed have data values aligned with those of the calibration dataset, although statistically not drawn from the same population.  In most cases the calibration dataset has a wider distribution than the XO-3b dataset, which makes sense as it has about 5.5 time the amount of data as XO-3b, and was designed to encompass a wide range of feature space so that it could address all science observations. Temperature values (denoted as T\_cernox) are quantized at the 1-2 mK level due to the digital bit resolution of the sensors.

 We compare also the observational programs of the calibration and science datasets.  The XO-3b dataset is extremely typical of exoplanet datasets in the archive, and follows a set of best practices suggested on the IRAC website\footnote{\url{https://irachpp.spitzer.caltech.edu/page/Obs\%20Planning}}.   Both targets have signal to noise ratios well within the linear regime of the detector.  This XO-3b dataset is slightly different from the calibration dataset in that it has a different frame time (2s on science data vs. 0.1s on calibration data). Frame time is not the important feature, instead well depth is important so having all data taken in the linear regime of the detector is important.  Data taken at longer frame times (greater than 2 seconds) might experience smearing of the systematic as the target moves around on the pixel within a single frame. The duration of the XO-3b observations is 8.5 hours.  Archival observations span the range from about 2.5 hours to days with the majority around 6 - 10 hours for an eclipse or transit.  The calibration observations were taken with 30 minute durations, which at 0.1s frame times is about 18000 images per observation.  For XO-3b at 8.5 hours in duration and 2s frame times, that is 15300 images per observation, a similar number to the calibration dataset.    

\section{Machine Learning} 
\label{sec:ML}
Here we briefly describe the ML algorithms used along with various modifications or improvements tailored to this project.  

We use a train/test split of $75/25$ which means that in all cases we train the algorithms on $75\%$ of the calibration data, saving $25\%$ on which the models were not allowed to train, allowing us to evaluate performance.  These test sets are randomly chosen from the calibration data. We confirm that for a single split, the test and train sets statistically cover the same ranges of feature space.  The median Anderson Darling p-value is 0.64 indicating that indeed the test and train datasets are drawn from the same distributions. We assume this holds for all splits as these splits are done many times internally to the process of building machine learning models. Decision trees and Random forests were built in python using scikit-learn \citep{scikit-learn} and XGBoost \citep{2016arXiv160302754C}.   We train all models on the training dataset and evaluate their performance on the test dataset.  To compare the many different models which result from using different hyperparameters, or different models (DT vs XGBoost), or using different feature sets, we use the XO-3b dataset.

\subsection{Decision Tree} 
\label{sec:DT}

Decision trees are most commonly thought of in terms of one dimensional classification problems where the outcome is to classify an object based on it's features. Classification means giving a discrete label to each item in a dataset (ie., is it an orange or an apple?) Regression implies solving problems which have a continuous distribution instead of discrete values. A one dimensional regression based on a decision tree looks like a linear approximation to the regression function.  We can extrapolate a single dimensional decision tree to a multi-dimensional space of features. The key to the machine learning aspect of decision trees is that the machine learns using the labelled data how to split up the feature space at each node of the tree to minimize a cost criterion.  There is no human intervention in deciding how to split the features.

Inside of the decision tree, a metric needs to be used to determine when and how to split each branch into subsequent branches.  This is done using mean squared error (MSE). MSE is the average of the squares of the deviations between observed and predicted values. Values closer to zero are better as they indicate lower deviations of the predicted values from their true values.  The advantage of taking the squares of the deviations (as opposed to the root mean square errors) is that it incorporates both the variance and bias of the predicted values. 

To build an initial model, we use our labeled dataset on our calibration target to build a single, basic decision tree with 20 features (xcen , xerr, ycen , yerr, flux\_unc, xfwhm, yfwhm, bg\_flux, sigma\_bg\_flux, noise pixel, t\_cernox, nine pixel values), and then use that tree to predict flux variations for the XO-3b dataset.  This tree took 18 minutes to train on a 6-core Intel Xeon E5 Mac desktop (hereafter referred to as Mac desktop).

\subsubsection{Decision Tree Hyperparameter Optimization}

Hyperparameters are parameters which are not used in the learning process itself, but which control the capacity of the model.  An example of a hyperparameter is the depth of the decision tree.  The depth is the maximum distance along the branches from root to the furthest leaf, or similarly the number of levels of branching.  This value must be set before training the tree, and should be set properly to prevent overfitting.  For example, allowing a model to have a very large number of branches might produce something which accurately reflects the training data, but would not be good at predicting values on data unseen.   It is therefore necessary to optimize hyperparameters, to accurately fit training sets while not overfitting training sets.

Hyperparameter tuning is an optimization task.  The most obvious way to do this is to perform a grid search over all reasonable hyperparameter values, choosing that hyperparameter set which optimizes some metric.  A full grid search is computationally very expensive, and we therefore employ a randomized grid search.  It has been shown by \cite{2012JMLR....13...1B} that a random search over hyperparameter space can be just as effective as a full grid search but take less time.  The idea is that taking 60 samples of any parameter space will find a maximum within 5\% of the true maximum 95\% of the time.  This assumes that the maximum covers at least 5\% of points in the grid.  We find this to be true because our decision tree and random forest models are not extremely sensitive to hyperparameter selection (see Section \ref{sec:xgboost}).

A further important detail in our hyperparameter optimization is that we use k-fold cross validation (CV).  At each random position in our hyperparameter space, instead of the machine learning a model on the entire training dataset, the training dataset is split into k different sections (folds). Then one fold is omitted from consideration and a model is learned on the remaining k-1 folds.  This proceeds k-1 times, with each of the folds being omitted once, and the remaining folds combined as the training set.  At the end of running all the folds, the average is taken of the accuracy of each fold to represent the overall accuracy of that position in hyperparameter space.  This technique of evaluating the model more than once at each position helps to estimate the predictive power that our model will have once it is used on real data while simultaneously limiting overfitting by holding out different portions of the dataset each time.  Note, this k-fold CV does not ever touch the actual test data that was set aside in the train/test split, so that at the end of our hyperparameter search, when we have discovered the ideal hyperparameters, we can then re-learn a model on the entire training dataset, and evaluate it on the test dataset, and the test dataset will not ever have been used in the learning process.  We use five-fold cross validation on our hyperparameter optimization.  k-fold cross validation is a very powerful tool that is widely used in many aspects of machine learning.

We search for the optimum values of the two free hyperparameters in the trees; the number of branches and the minimum number of leaves per branch.  A branch will not split if fewer than this minimum number of leaves remain in the resulting branch. This results in maximum depth of 34 branches and minimum leaves of 13 per branch.   To give the reader an idea of how sensitive the results are to optimizing hyperparameters, in this particular decision tree optimization, the mean test score increases linearly by a factor of three from max depth of one to max depth of about 15, then levels off through the complete range of tested max depths.  Adjacent points in a grid search provide test sample scores within a few percent of each other; therefore adding further justification to using a randomized search instead of a full grid search.  The same holds true for boosted random forests as discussed below.

\subsubsection{Scoring}
 As a way of scoring how well the trees are accomplishing their goal of fitting the intrapixel response function for each of the grid points in our hyperparameter optimization, we use both the scikit-learn built-in metric of mean squared error and our own scoring method.  Both scorers are used in the optimization such that at each grid point, the model is evaluated, and scores tabulated for each scorer.  Mean squared error(MSE)is the average of the squares of the deviations between observed and predicted values.  We found that some models which were optimized based on MSE actually had bad performance on the XO-3b light curves; either by removing the eclipse signature, or not removing the systematic, which lead us to design our own metric based on fitting injected transits.

 For our own scoring method we choose to measure the model's ability to recover the transit depth of an injected transit. We refer to this method as 'injected depth accuracy' (or 'acc' in Table \ref{table:compare}). We could conceivably also have used transit duration or timing, however depth seems to be used more frequently in scientific analyses.  Within the k-fold CV, we inject transits into the test fold.  Then, to score that fold, we correct the data with the model and try to recover the injected transit depth.  Because the test data is drawn from the same dataset as the training data, it intentionally does not have any astrophysical variations, and therefore no inherent transits, which leads to our injection strategy.  Each injected transit has a randomly chosen depth with the ratio of the planet radius to star radius in the range of 0.03 to 0.1. Injecting transits allows us to know the true depth of the transit for calculating this scorer.   Our definition of this scorer is taken from \cite{2016AJ....152...44I}, and is defined to be the ratio of the intrinsic to measured error.  

\[ acc \equiv \sigma\textsubscript{phot\_depth} / RMSE \]

where $\sigma\textsubscript{phot\_depth}$ is the photon noise variance in the transit depth and RMSE is the square root of the MSE.  See the above reference for a full derivation of the variances.

 Both Scoring metrics are used in the hyperparameter optimization.  After optimization, the best performing 'mse' and 'acc' models are then tested against XO-3b eclipse depths.  Only those models which do well at recovering the average XO-3b eclipse depth are discussed here and in table \ref{table:compare}.

%\subsubsection{Feature Importance}
%When evaluating decision trees, we can make an estimate of feature importance, which tells us how often the features are used in decision nodes. A feature which is used more often to make key decisions in the tree will have a higher feature importance. We use this feature importance parameter to remove features which have the smallest impact on the results, and those are temperature, and position covariance.   Those two features are less important by at least two orders of magnitude than the most important features.

\subsubsection{Model Performance}
We compare our optimized decision tree model with other published methods of reducing IRAC data.  The best way to test the ML models is to use them on real data for which we think we know the ground truth of what the eclipse depths are.  This will be the most useful metric since it is how the methods will actually be used in practice.  We do this for the set of ten archival observations of XO-3b which have previously been analyzed in \citep{2015ApJ...811..122W, 2016AJ....152...44I}and are described in Section \ref{sec:data}.  Fitting the model to the XO-3b dataset takes mere seconds.  We compare our median measured eclipse depths to average values from the seven other teams.  

 Fitting the eclipses is an important step in this process which introduces differences between literature values on the same dataset.  We choose to use Levenberg–Marquardt fitting on a BATMAN model \citep{2015PASP..127.1161K} including fitting the out of eclipse light curve to have slope and curvature.  We find this gives the best fits to the eclipses and is a physically motivated, model based, fitting.  We choose the simplest possible fitting scenario where all parameters except the eclipse depth and the out of eclipse light curve intercept, slope and curvature are fixed to the current literature values.  Those fixed parameters are period, central time of the transit, inclination, semi-major axis, planet radius, transit depth, eccentricity, and argument of periastron. We do not fit limb darkening parameters because these are secondary eclipses and infrared data. Fixing most parameters works the best since we only have a small fraction of the total light curve.   We allow the minimum fit eclipse depth to be 1100ppm, since we know from the literature what to expect for this planet, and we do not want to allow un-physical eclipse depths.  This results in some eclipse depth fits bottoming out at 1100ppm which is really an indication that the fitting function has failed.  Statistically speaking this 1100ppm value is a floor.  Because of the use of a floor in our dataset, we measure median values of the ten XO-3b observations instead of a mean.  Median values will more accurately represent the peak of the distribution without making assumptions about the shape of the distribution.  In all cases, the measured median values are well within one standard deviation of the measured mean values when not including the floor values.  Standard deviations are calculated as expected including the floor values. The number of observations where the floor value is the best fit are listed for each model in Table \ref{table:compare} 

The average eclipse depth of all other seven teams reduction is $1520 +- 102$ppm.  Our decision tree model finds a median eclipse depth of $1530 +- 290$ppm.  This is promising, but the large spread in eclipse depths is concerning.  We seek to improve this initial model by switching models from a decision tree to a random forest and by manipulating our feature set.

\begin{table*}[]
\caption{Comparison of model parameters for all models discussed in this paper}

\hskip-2.0cm\begin{tabular}{lllllllllll}
\hline
Model  & num  & scorer & max & min smpls & num  & eta   & min child & median  & stddev. &num \\
type & features &   & depth & leaf & estimator  &  & weight & ecl. depth & depth & floor \\
\hline \hline
DT         & 20          & mse    & 34         & 13                 &               &       &                    & 1530              & 290      & 0    \\
XGBoost    & 16          & acc    & 3          &                    & 1252          & 0.2   & 7                  & 1459              & 200      & 1    \\
XGBoost    & 9           & mse    & 3          &                    & 1252          & 0.1   & 7                  & 1473              & 209      & 2    \\
XGBoost    & 14          & mse    & 6          &                    & 5591          & 0.01  & 11                 & 1100              & 9        & 9    \\
XGBoost    & 20          & mse    & 4          &                    & 5631          & 0.005 & 3                  & 1413              & 330      & 1    \\
XGBoost    & 11          & mse    & 4          &                    & 5631          & 0.005 & 3                  & 1452              & 297      & 1   \\
\hline
\end{tabular}

\label{table:compare}
\end{table*}

\subsection{Random Forests} 
\label{sec:RF}

 A random forest is an ensemble of decision trees. In the case of a random regression forest, many trees are fit and the mean of the output for each input set is used as the final value.  There are two things that make each tree in the forest different from other trees which gives rise to the "random" on the name of this method. Each tree is built from a random subset of the data, and at each decision node in the tree, a random subset of the features are considered in splitting the data.  Therefore each tree is slightly different, and the forest ensemble averages, hopefully, to a better overall regression fit.  

The advantage of a forest of random trees is that it minimizes over-fitting since each tree only sees a random subset of the data with a random subset of features at each node.  One disadvantage of regression forests (and decision trees) is that the trained model poorly predicts values outside of the range present in the training set.  We emphasize again that the models derived in this paper only work for data taken at the sweet spot of the detector.

\subsection{Boosted Random Forests}
\label{sec:xgboost}

\begin{figure}
 \centering
  \includegraphics[width=0.5\textwidth]{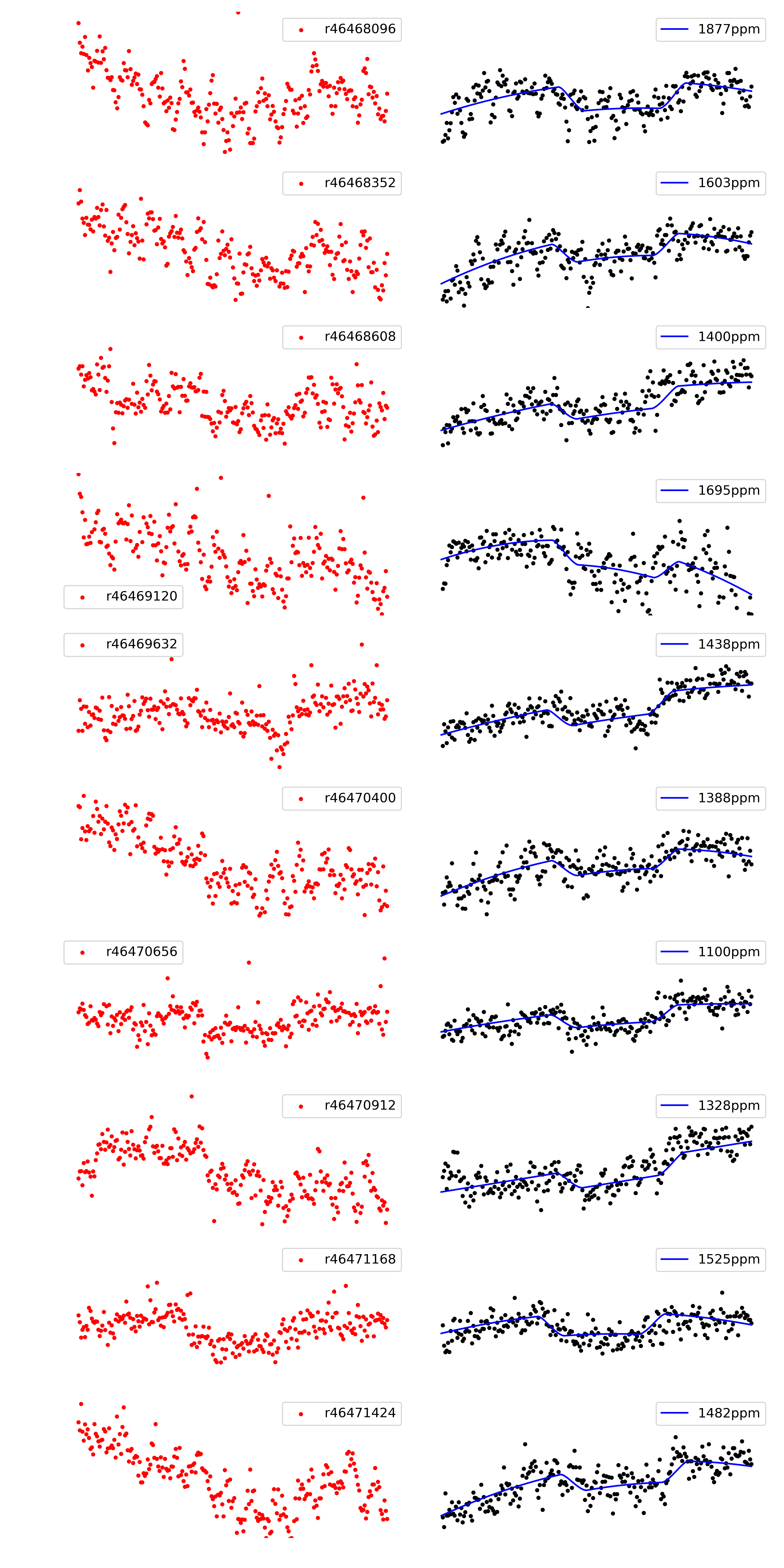}
 \caption{\label{fig:light_curve_fits_16} Raw (red) and XGBoost corrected (black) light curves for the ten observations of XO-3b using the 16 feature model.  Data is binned for display purposes.  Best fit eclipses are shown in blue and are the source of eclipse depth plotted in later figures.  Observation names are listed on the raw light curves.}
 \end{figure}

Random forest models using scikit-learn's randomforestregressor model are the slowest to run (taking over a week on a Mac desktop), and so we have not tested many of these models, instead choosing to work with faster, boosted random forests.

Boosting is a way of weighting the input values to the trees such that those values which are not as well fit are given higher weight in future trees.  In this way, as more trees are added to the forest, the trees hopefully get smarter.  We use XGBoost for this work because it is faster than than scikit-learn's GradientBoostingRegressor.

We remove four features from our feature set for further model training  (x \& y centroid, cernox temperature, and noise pixel). We choose not to include x and y centroid uncertainties.  After preliminary testing with those features included, we find that although metrics are generally improved with inclusion of centroid uncertainties, the fits to the real data make light curves which show greater variability than the raw light curves.  Centroid errors have a measure of the flux in them since we use light weighted centroiding, so the worse performance is probably due to overfitting when including a feature which is correlated with results.

We also do not include the temperature in our feature set. Although it seems plausible that the pixel response might be correlated with array temperature, we find no evidence of any correlation at the small level of variation in t\_cernox that exists (the IRAC array temperature is maintained within a narrow range by heaters in a feedback circuit).  In fact, when we include t\_cernox in the model training, the variation in eclipse depth from epoch to epoch increases unrealistically.

%While it seems promising that temperature of the arrays would be correlated with intrapixel response, we find no evidence that this is an important feature in testing models, with it only increasing the spread in eclipse depths measured when included in the model training.  

Lastly, we do not include noise pixel because XFWHM and YFWHM are the individual, more specific, components of noise pixel and redundancies could lead to the model overfitting the data.

We now train and optimize models with the remaining 16 features (xcen, ycen, flux\_unc, xfwhm, yfhwm, bg\_flux, bg\_unc, and nine pixel values)

\subsubsection{XGBoost Hyperparameter Optimization}

The hyperparameters we vary to control the flexibility of the model are the number of trees in the forest, the depth of the trees (number of branches), the number of samples at each end node (leaf), and learning rate  (also called eta). Learning rate is a way of slowing down the boosting that occurs with every new additional tree to prevent overfitting. For boosting, it is also important that each tree not be allowed to grow so large that it could overfit the data.  As such, we keep our trees pruned, searching over maximum depths in the range of three to twenty levels per tree.  As with the decision trees, we employ a randomized search with five fold cross validation.  The resulting best fit parameters are 1252 trees, tree depth is 3, number of samples at nodes is 7 and learning rate is 0.2.  This was found using the transit injection accuracy as the scorer.

Hyperparameter search with XGBoost on 16 parameters took 35 hours on a Mac desktop; with all 20 parameters it took 45 hours.

\subsubsection{Model Performance and Attempts at Improvement}
\begin{figure*}
 \centering
  \includegraphics[width=0.65\textwidth ]{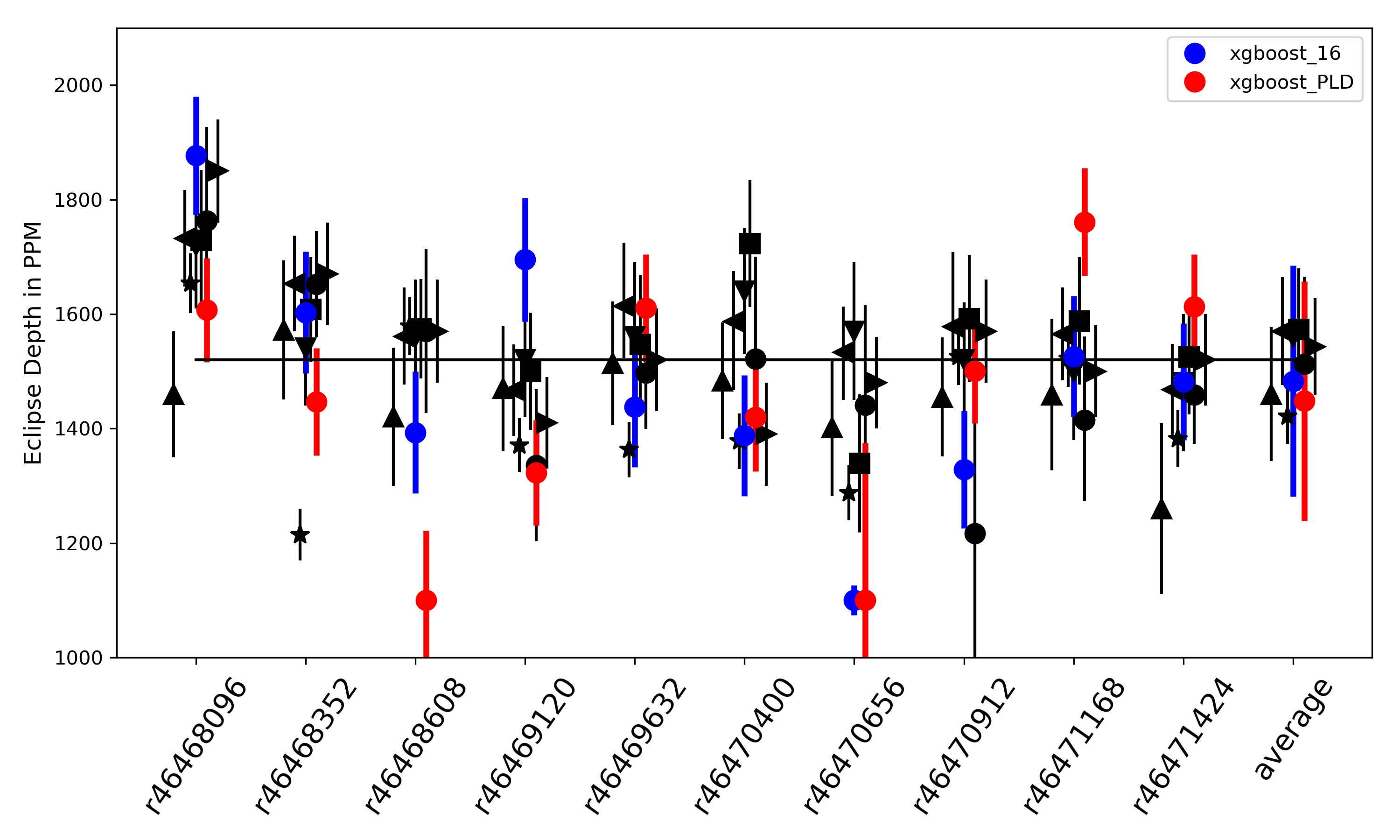}
 \caption{\label{fig:eclipse_depth} XGBoost model comparisons using the two different feature sets. Blue shows the 16 feature model and red shows the nine feature model which emulates the PLD method by using only the nine pixel values as features. Eclipse depth is shown on the y-axis as a function of the ten observations and the average of all ten.  Black points are from \citet{2016AJ....152...44I} and represent the other seven methods of systematics removal.}
 \end{figure*}

Our best XGBoost model finds a median eclipse depth of $1459 \pm 200$ppm.  Figure \ref{fig:light_curve_fits_16} shows the ten raw and corrected XO-3b light curves as corrected with the best XGBoost model with 16 features.  Legends show the fitted eclipse depth for each observation.  Figure \ref{fig:eclipse_depth} shows the eclipse depths of the XGBoost model in comparison to those from the literature for the XO-3b dataset. 

 Figure \ref{fig:eclipse_depth} shows some interesting correlations between the different models in the literature and this model. For example the first two observations show a deeper than average eclipse depth measurement from all literature methods.  The fourth observation shows literature values all having lower than average eclipse depth measurements.  Our XGBoost model seems to not follow the correlations among the other methods - or at least not as strongly.  As discussed below in Section \ref{sec:truth}, this could be an indication of eclipse depth variation due to astrophysical sources.  On the other hand, this might be indicating that there is something in common to the other methods which produces these correlated depth measurements. 

Literature average eclipse depths, uncertainties, and main authors are listed in table \ref{table:avg_depth} along with the results from this work.  This table is copied largely from \cite{2016AJ....152...44I} which summarized the data challenge run at the 2015 American Astronomical Society meeting.  Author column represents those authors who worked on the XO-3b dataset and are co-authors on the data challenge paper.  Refer to that paper for further details on the methods and examples of individual light curve fits.

\begin{table*}[]
\caption{Comparison of average eclipse depths and uncertainties for ten XO-3b archival observations}
\begin{tabular}{lllll}
\hline \hline
Method           & Avg. Eclipse  & Unc. & Plot               & Author    \\ 
                 & Depth (ppm)         & (ppm)            & Symbol             &      \\[0.5ex]    
\hline
BLISS            & 1543              & 85          & $\blacktriangleright$     & Stevenson \\
GP               & 1513              & 152         & \textbullet         & Evans     \\
ICA              & 1560              & 111         & $\blacktriangledown$   & Morello   \\
KR/Data          & 1570              & 94          & $\blacktriangleleft$    & Wong      \\
KR/Pmap          & 1460              & 117         & $\blacktriangle$      & Krick     \\
PLD              & 1573              & 107         & $\blacksquare$    & Deming    \\
SP(K2)           & 1421              & 48          & $\star$ & Buzasi    \\
Average          & 1520              & 102         & \textemdash      &           \\
XGBoost(16) & 1459              & 200         & \tikz\draw[blue,fill=blue] (0,0) circle(.5ex);     & this work \\
XGBoost(9)  & 1473              & 209         & \tikz\draw[red,fill=red] (0,0) circle(.5ex);        & this work \\ [1.0ex]
\hline
\end{tabular}
\label{table:avg_depth}
\end{table*}

After running the hyperparameter optimization on the feature set with 16 features we went back and tested training many different models with various feature set changes, normalization changes, scorer changes, and model changes.  These are all listed in table \ref{table:compare}. We tried four different feature sets ; 1) removing the flux\_unc and background error features from the standard feature set, leaving 14 features, 2) using all 20 features,  3) emulating pixel-level decorrelation  \citep[PLD][]{2015ApJ...805..132D} by using only the nine pixel value features, and 4) remove this pixel values as features and using the remaining 11 features.  For each of these cases, we re-ran the random search with CV to search for the best hyperparameters and then applied the results to the XO-3b dataset.  

We chose the first case of removing the flux\_unc and background error values from the feature set because those were determined to be the most important in the feature set.  We wanted to test what would happen if we removed the most important features and forced the model to work with the more standard features used in the literature of flux location and pixel value.  The results of this were models which routinely removed signal from the light curves resulting in eclipse depths which were not well fit ( 9 of the 10 observations had best fits at the floor value).

Secondly we chose to include all features, even those that had been eliminated early on in this work so that we could be sure that after doing some optimization, we hadn't missed any important features.  This turns out not to be the case, the XO-3b light curves corrected with models built on all 20 features are very noisy.  The XGBoost model has $1413\pm 330$ppm.

\begin{figure}
 \centering
  \includegraphics[width=0.5\textwidth]{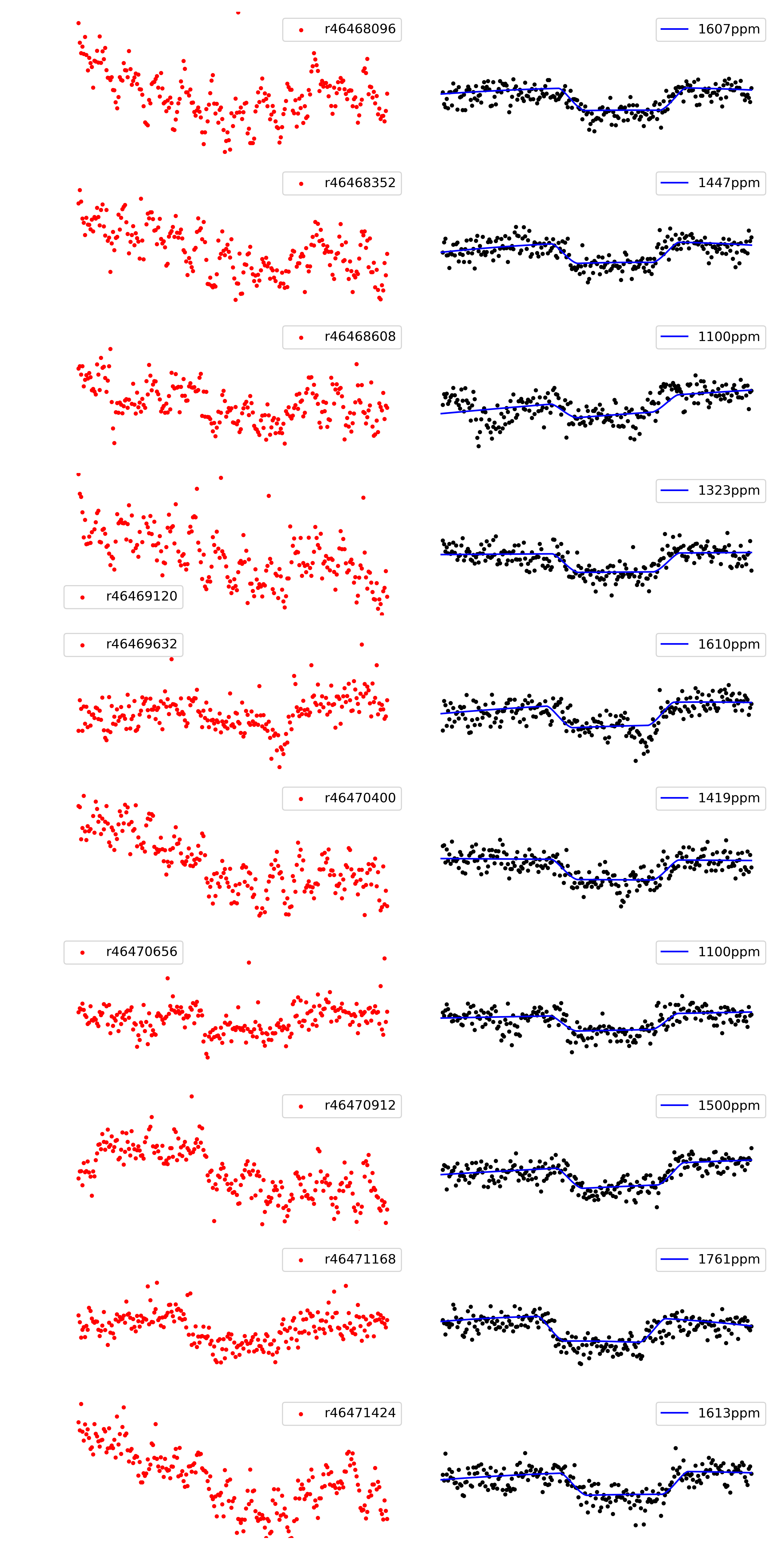}
 \caption{ \label{fig:light_curve_fits_9} Raw (red) and XGBoost corrected (black) light curves for the ten observations of XO-3b using only the nine pixel values.  Same as Figure \ref{fig:light_curve_fits_16}}
 \end{figure}

Thirdly, following the success of PLD, we wanted to check the performance of a boosted forest with just the nine central pixel values.  As seen in Figure \ref{fig:eclipse_depth}, the PLD-like model does give similarly good results as the 16 feature model.  In this case we normalized the pixel values by their sums to stay in kind with PLD. Normalizing by the sum of all the pixels should effectively remove the astrophysical signal from the pixel values, which means that our central pixel feature is not highly correlated with the final flux. This XGBoost model has a median eclipse depth of $1473 \pm 209$ppm.  Interestingly, the corrected light curves from this model shown in Figure \ref{fig:light_curve_fits_9} look better than the full 16 model, but there are now two observations with an essentially unfit eclipse depth (recorded as 1100 ppm), compared to one floor value for the full 16 feature model.

Finally, wondering if the pixel values were overly correlated (even when normalized), we ran a boosted forest model with all 11 features which were not the pixel values.  Despite the limited number of features, this model took 29 hours to learn.  This XGBoost model has a median eclipse depth of $1452 \pm 297$ppm.  Six of the ten light curves are well corrected and have nice fits, but the other four are not well fit, leading to the unacceptably large scatter.  

%Testing a PLD like model also led us to some feature engineering.  We changed the normalization on the pixel values to be the sum of all pixel values (instead of the mean) and background subtracted the pixel values. Background values are taken to be the average of the outer ring of pixel values from the seven by seven pixel box originally sampled from the image.  This is in kind with \citet{2015ApJ...805..132D}.  Background subtraction does not make a big difference, however

%\begin{figure}
% \centering
%  \includegraphics[width=0.5\textwidth]{XO3_ML_compare_sumnorm_9_xgboost.png}
% \caption{\label{fig:eclipse_depth_9} XGBoost model comparison with only the nine pixel values as features}
% \end{figure}

\section{Discussion}
\label{sec:discussion}

Now that we have a functional XGBoost model which we have chosen based on a rigorous hyperparameter search, we would like to be able to understand what this model is telling us about our data. As we are using a random forest with many thousands of trees, we need a way of interpreting and visualizing the model and its output.  Interpreting the model is important for both understanding which features are causing the systematics, and for sanity checks of knowing if the model is finding linear correlations in features instead of actually modelling the systematic.  

 We test the viability of standard methods of interpreting our models, all with limited results, discussed in detail in the appendices.  We test conditional expectation plotting, feature importances, and permutation importances. Conditional expectation plots showed nothing of value in interpreting the models.

Feature importance captures a gain value for the number of nodes in the tree which use that particular feature (assuming that if a node uses a feature, then it is important).  There are three methods for calculating feature importance, and after employing all three, we find contradictory results among the methods.  Permutation importance is randomly permuting the values of each feature and measuring how much that permutation negatively impacts the scoring metric.  We find that the permutation importances show again different results than the feature importances, and reveal values of importance which are not very high.  The likely interpretation of both the feature and permutation importances is that the features are correlated. However, running correlation tests and removing weakly correlated features from the model fits results in models which do not perform as well as the original models on measuring XO-3b eclipse depths.  See the appendices for further examination of model interpretation methods.

\subsection{Ground Truth}
\label{sec:truth}

We do not actually know what the true eclipse depth is for XO-3b.  We can postulate that the average of the other seven data reduction methods gives the correct true eclipse depth, but we do not know that for sure.  For this reason, the 2015 data challenge included simulated data in addition to the XO-3b dataset.  That simulated dataset was generated by the Spitzer Science Center so that we knew the true eclipse depth and all the positions/pixel values/ fluxes etc., that went into building the dataset.  However, this simulated dataset could not possible include an accurate simulated training set including all the features that we currently are using for this ML forest application, so we cannot test our XGBoost models on that simulated dataset.

In addition to not knowing the true eclipse depth, we also do not know the true variation in eclipse depth. The 200ppm standard deviation($13\%$) that we measure on the ten eclipse observations is likely part systematics and potentially part real temperature variation between orbits.  \citet{2014ApJ...794..134W} finds this variation of order $5\%$ from the mean to be consistent with measurement uncertainty indicating that condensation or turbulent mixing are not significant effects for the XO-3b atmosphere.

\subsection{Catastrophic Errors}
\label{sec:floor}
 As can be seen in the "num floor" column of Table \ref{table:compare}, some models produce data for XO-3b for which our fitting function can not find a reasonable eclipse depth (labeled as hitting an 1100ppm floor or minimum value in the fitting function).  These can be considered catastrophic errors of either the model or the fitting function's ability to handle the input data.  Note that only the DT model has no fitted eclipse depths at this floor level.  This leads us to conclude that a) careful consideration of prior constraints is important, b) datasets of this nature with multiple observations of the same target are invaluable in vetting systematics removal models, and c) potentially more sophisticated and more uniform fitting techniques are warranted.

%It is interesting that for xgboost with 16 parameters that xFWHM and yFWHM are not higher in importance when our experience tells us that these parameters effect the systmatics - XXX do I have plot sof NP vs. amplitude/strength oor duration of the sawtooth?

%why are flux\_unc and bkgerr so important? XXX maybe this is a failure of the model that it chooses these features so often in the decision nodes?

%MSE and r2 scoring methods give roughly the same results in terms of average eclipse depth and standard deviation as the injected transit accuracy which we developed despite often having significantly different hyperparameters.  This is potentially a further indication that hyperparameter optimization only provides minimal gains in terms of model performance.

%Interesting that 9 pixel PLD like model gives similar results to the 16 feature models even though feature importance for the 16 feature models has the highest imporantce not going to the pixel values.  Not sure what this is telling me.  XXX.

%We tested flux\_unc and bg\_unc from our feature list (keeping the remaining 14 features).  These two features have different enough distributions in Figure \ref{fig:dist} that we worried the model would find errant solutions when including those features.  In fact, the model does a better job when those two features are included.  The average eclipse depth without them is 1100ppm.  In other words, removing those two features results in the undesired effect of removing the eclipse from the light curve.  

\section{Conclusions} 
\label{sec:conclusion}

Using ML, we have made a model to reduce systematics in IRAC light curve observations.  We benchmark this model against other literature methods for systematics reduction using the archival dataset of ten eclipse observations of XO-3b. We measure the median eclipse depth of XO-3b to be $1459 \pm 200$ parts per million.  This is comparable to the average depth in the literature, however the spread in measurements is something like 30-100\% larger than those literature values, depending on the reduction method.  This work also presents the first attempt at feature analysis of intrapixel response for Spitzer photometry.

This paper serves mainly as a cautionary tale and pathfinder for anyone else attempting this type of application of ML to systematics reduction.  Because we had such promising initial results, we put a lot of effort into feature engineering and hyperparameter optimization in hopes of finding a better model.  The lessons we learned are the following.

Some models can make great looking light curves that underestimate the eclipse depth if we take the true depth to be the average of all the other methods.  This can be misleading.  All new systematics reduction methods should compare their results against the results of the XO-3b dataset.  This is why we initially included injecting transits as an accuracy score.  Also, for future instruments and future telescopes, it is invaluable to have a test dataset which all can reduce and compare results.   Especially useful has been the test dataset with multiple observations of the same target to test the reliability of the methods in light of catastrophic failures.

The nine pixel model gives similarly good results as the full 16 feature model, giving credence to the PLD method \citep{2015ApJ...805..132D}.  This work also implies that there is no smoking gun feature in our feature set which magically produces accurate models.  We conclude then that researchers have been on the right track with using features included here to remove systematics. 

 We test decision tree models as well. Although the resulting optimized DT model has an uncomfortably high standard deviation amongst the 10 XO-3b observations, it is the only model with no failures where failure is defined as not finding a physically reasonable eclipse depth.

 Prior to beginning this work, it was not apparent how sensitive the eclipse depths are to the fitting method or fitting function chosen.  Caution and care should be taken when choosing how to do fitting for astrophysical parameters.

We cannot use time as a feature in our models because then we would not be able to use the model for science observations taken outside the time range for which we have calibration data.  It could be that self calibration methods work on IRAC data because there is a time-varying component to the intrapixel response function.  We are unable to comment on that possibility with this work. 

We still believe that using a calibration dataset to train ML models is a viable way of removing noise from astronomical datasets.  In future works we intend to use deep learning, meta ensembles, or stacking models to search for improvement for Spitzer IRAC data.

It is important to continue pushing the envelope on methods that can independently, quickly, without hand-holding, uniformly reduce astronomical data so that we can analyze ensembles of sources.  Science that is enabled by uniform reduction of ensembles of sources includes looking for correlations between astrophysical parameters and energy transport efficiency to better understand the influences of various system properties, such as stellar and planetary mass, stellar metallicity, stellar rotation speed, planet day side surface temperature (irradiation), planet atmosphere chemical composition and vertical pressure profiles, location of the primary hot spot, angle between stellar rotation axis and planetary orbit axis, etc. This is particularly true for the rich Spitzer IRAC exoplanet archive, but also generally true for the coming deluge of astronomical data from ground and space-based telescopes.

We make our code and trained models publicly available on github.  \url{https://github.com/jkrick/XGBoost_IRAC}.

\acknowledgments
We thank the anonymous referee for their time and care in providong very useful comments on this manuscript.  This work is based [in part] on observations made with the Spitzer Space Telescope, which is operated by the Jet Propulsion Laboratory, California Institute of Technology under a contract with NASA. This research has made use of NASA's Astrophysics Data System. This research has made use of the NASA/IPAC Infrared Science Archive, which is operated by the Jet Propulsion Laboratory, California Institute of Technology, under contract with the National Aeronautics and Space Administration. The acknowledgements were compiled using the Astronomy Acknowledgement Generator. This research made use of scikit-learn \citep{scikit-learn}.

\bibliography{planets.bib}
\facility{Spitzer (IRAC)}

\appendix
\section{Feature Importance}
 Feature importance captures a gain value for the number of nodes in the tree which use that particular feature (assuming that if a node uses a feature, then it is important).  There are three ways of calculating feature importances. The 'gain' method calculates the improvement in accuracy that feature generates at each branch splitting.   The 'coverage' method calculates the relative number of times a feature is used in the final node of all trees.  The 'weight' method is a percentage weight of the number of times that feature occurs in the model.  The gain method is the most commonly used algorithm.  

\begin{figure*}
 \centering
  \includegraphics[width=0.8\textwidth]{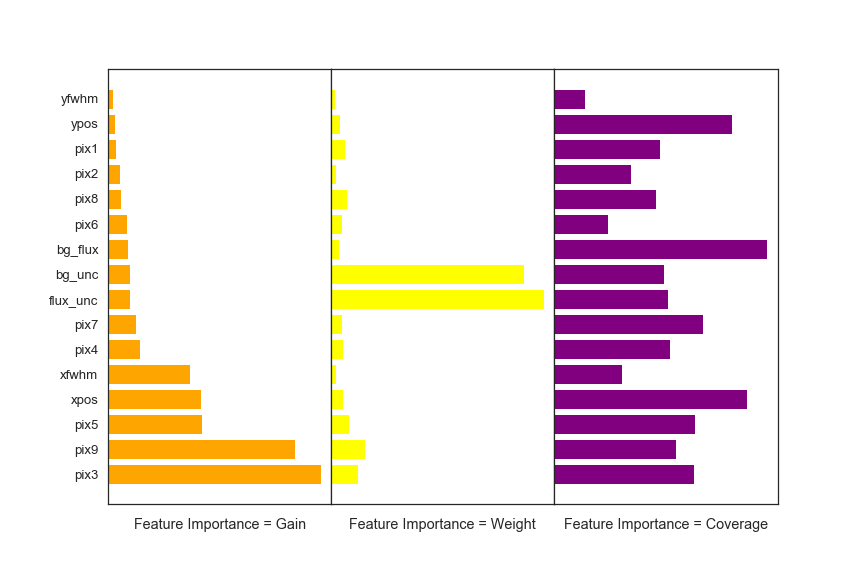}
 \caption{\label{fig:FI}{\bf Feature Importances} shown for three different methods.  Note the discord amongst the methods.}
 \end{figure*}

We use the Eli5 python module to calculate feature importances.  Figure \ref{fig:FI} shows the three feature importances for our best fit XGBoost model.  In the leftmost gain plot, the importance of pixels 3 and 9 above all other features is interesting as they are both on one side of the 9 pixel box we use for features (counting three pixels along the x axis, then increase y value by 1 and count another 3, etc.  In this scheme pixel 5 is the center pixel).  This is consistent with xcen and xfwhm also being at the top of the list of important features.  Interestingly ycen and yfwhm are relatively unimportant. We do know that the peak of the pixel response is to the 'right' of center, ie., on the side of the pixel towards pixel 3 and 9 such that when the star moves left or right, pixels 3 and 9 will experience larger variations.  But, so would pixel 6, which is not indicated here.  Potentially this feature importance metric is telling us that shifts in the x direction are accompanied by shifts in the y direction.  We speculate that shifts in the x direction are more important in predicting the IRAC systematics.  

There are no known X-only motions in the telescope.  The former saw-tooth pattern seen in many exoplanet observations was in both x and y directions and is due to thermo-mechanical expansions and contractions caused by a periodic heating cycle within the spacecraft.  There is a known long term drift in the y direction.  This drift is known to be due to an inconsistency in the way that velocity aberration corrections are handled by the spacecraft's Command and Data Handling computer (C\&DH) and the star trackers.

Unfortunately the three feature importance methods give us contradictory results, with the gain showing that pixel 3 and 9 (the bottom and the top corner) are the most important features, whereas the weight method implies that 'flux\_unc' and 'bg\_unc' are the most important.  The 'coverage' method shows that again different features are the most important, although none are highly favored. This is a known problem of feature importances. \cite{Strobl2007} shows through simulations that feature importances are often unreliable when data have varying scales or some data have multiple observations with different outcomes.

\section{Permutation Importance}
\begin{figure}
 \centering
  \includegraphics[width=0.5\textwidth]{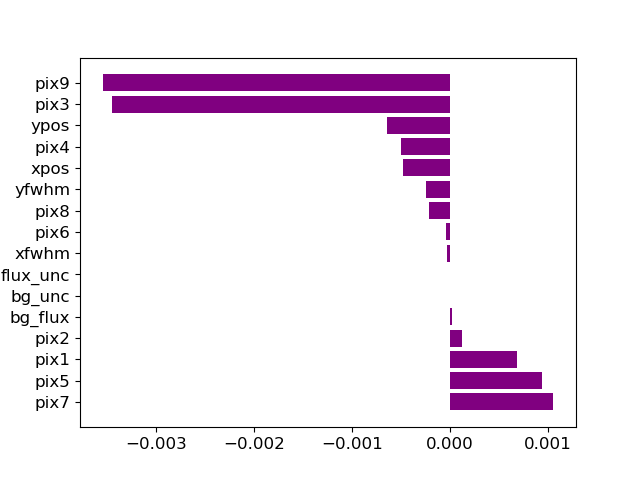}
 \caption{\label{fig:PI}{\bf Permutation Importances} implying that pixel 5 is the most important feature and most other features are not important to modelling the intrapixel response function}
 \end{figure}

The unreliability of feature importance leads us to look at permutation importances. The basic idea of permutation importance is to randomly permute the values of each feature and measure how much that permutation negatively impacts the scoring metric. In this case we use $R^2$ as a metric which is the proportion of the variance that is predictable from a model. We first calculate the $R^2$ value of the best fit model. Then, we shuffle the values of one feature so that feature now has essentially random values (but with the correct scaling), and re-calculate the $R^2$ metric. The permutation importance is the size of the decrease in value of $R^2$ calculated on the base model and on the shuffled model. If we shuffle the values of a feature, and it doesn't impact the $R^2$ value(zero permutation importance) or increases the $R^2$ value (negative permutation importance), then that feature must not be very important.  If randomizing the order of values of a feature decreases $R^2$ (positive permutation importance), then the feature makes a measurable contribution to the model.  We run 50 iterations of every feature being shuffled to assure random distributions. 
 
Permutation importances are shown in Figure \ref{fig:PI}. Negative permutation importances imply the model improved (albeit slightly) with randomizing that feature; or that those features are unimportant.  Looking at the results, it is easy to imagine a model where the pixel values are important since they contain information about fractional flux and therefore about the more exact position of the star on the pixel.  It is strange that permutation importance shows the exact opposite of feature importance in putting pixels 3 and 9 at the absolute bottom of the list. Interestingly, bg\_unc and flux\_unc are sort of agnostic, shuffling their values reveals that they are not important and do not change the results.  Another interesting result is that xcen and ycen are also agnostic and apparently not very important in determining the accuracy of this model.

The test set permutation importances are exactly the same as the training set ones which gives us confidence that we are not overfitting.

Because none of the permutation importances are high in value, we are concerned that some of the features are correlated.  If features are correlated, then permuting one of them will not have a large effect on the model performance because the model can use information from the correlated feature. 

\begin{figure*}
 \centering
  \includegraphics[width=0.8\textwidth]{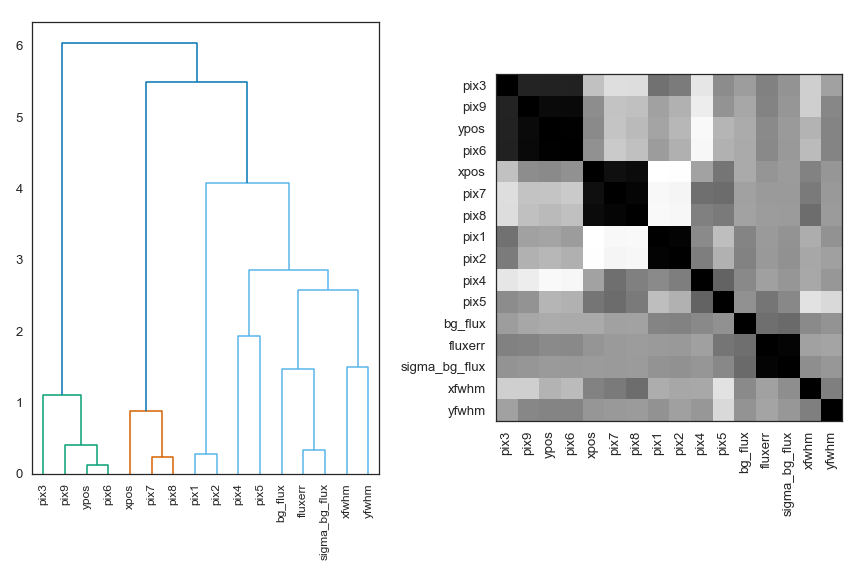}
 \caption{\label{fig:spearman}{\bf Spearman Rank Correlation Tests} The left plot is a dendrogram where the vertical height of the lines connecting clustered objects on the x-axis shows the extent to which those features are correlated  where longer lines imply weaker correlations. The right plot is a greyscale heat map where darker colors imply stronger correlations.}
 \end{figure*}

We test for correlations among features using Spearman rank test, shown in Figure \ref{fig:spearman}.  The left plot is a dendrogram where features that are correlated are clumped together at the bottom of the plot. The vertical height of the connecting lines shows how well features are correlated;  longer lines implies weaker correlation. This same information is shown in a heat map on the right where higher correlation coefficients are shown in lighter colors, so for example, pixel 3, pixel 9, ycen , and pixel 6 appear correlated; as do xcen with pixels 7 and 8 (pixels seven and eight are connected in the x direction).  The uncertainties in the flux and background flux also appear correlated, as are some adjacent pixels.

Using this information, We repeated a random search with CV on XGBoost models with the ten indicated features that are uncorrelated ('xcen ', 'ycen ', 'flux\_unc', 'xfwhm', 'yfwhm', 'bg\_flux',  'pix1',  'pix3', 'pix4', 'pix5').  After finding the best fit hyperparameters, we then tested that resulting model on the XO-3b dataset.  The search took 59 hours on a Mac desktop. The  median eclipse depth of that model is  $1175 \pm 180$ppm, with 3 observations having the minimum fitted eclipse depth.  This is much smaller than the literature value, indicating that removing these features decreases the adequacy of the model to estimate systematics in the XO-3b eclipse data, regardless of what importance tests are showing.  The full feature set is apparently preferred.

\end{document}